\documentclass[a4paper,11pt]{article}
\usepackage{pos}
\usepackage{hyperref}

\title{Pseudoscalar Screening Mass at Finite Temperature and Magnetic
Field from Lattice QCD with Physical Quark Masses}
\ShortTitle{Pseudoscalar Screening Mass at finite $T$ and $eB$}

\author*{Rishabh Thakkar}
\author{Heng-Tong Ding}
\author{Jin-Biao Gu}
\author{Sheng-Tai Li}

\affiliation{Key Laboratory of Quark and Lepton Physics (MOE) and Institute of Particle Physics,\\
Central China Normal University, Wuhan 430079, China}
\emailAdd{rishabh@ccnu.edu.cn}
\abstract{Understanding the screening mass of pseudoscalar mesons at finite temperature and magnetic field
is crucial for comprehending the behavior of strongly interacting matter under extreme conditions,
such as those found in the early universe or inside neutron stars. Additionally, in heavy ion collisions,
strong magnetic fields are generated, which could significantly influence the properties of the quark-gluon plasma. The study of these screening masses provides insight into the modifications of mesonic
properties in such environments, which is essential for the theoretical understanding of Quantum
Chromodynamics (QCD) phase transitions and the properties of the quark-gluon plasma.
Here, we present continuum estimated lattice QCD results on the screening mass of neutral pseudoscalar
mesons at finite temperatures and nonzero magnetic fields. The simulations used (2+1)-flavor lattice
QCD simulations using physical quark masses employing the HISQ/tree action. The continuum
estimation was carried out using lattices having temporal extents $N_\tau$ = 8, 12, and 16, all having aspect
ratio $N_\sigma/N_\tau$ = 4. The investigated temperature ranges from 145 MeV to 166 MeV, while
the magnetic field strength varies from 0 to 1 GeV$^2$. We discuss the dependence of the screening masses of
various neutral pseudoscalar mesons on temperature, magnetic field strength, and quark mass.}

\FullConference{The 41st International Symposium on Lattice Field Theory (LATTICE2024)\\
 28 July - 3 August 2024\\
Liverpool, UK\\
}

\begin{document}

\maketitle

\section{Introduction}
Magnetic fields play a crucial role in shaping the properties of QCD matter under extreme conditions, such as those encountered in high-energy heavy-ion collisions, inside neutron stars, and during the early universe \cite{endrodi2024qcd}. The interplay between magnetic fields and the strong interaction provides valuable insights into the QCD phase structure and the behavior of strongly interacting matter.

One of the significant effects of magnetic fields on QCD matter is the suppression of the pseudocritical temperature $T_{pc}$ with increasing magnetic field strength \cite{bali2012qcd}. This suppression is accompanied by a phenomenon, known as inverse magnetic catalysis (IMC) \cite{bali2012qcd1}, where the light quark chiral condensate, which is the order parameter of the chiral symmetry, decreases with increasing magnetic field. This contrasts with the previously expected magnetic catalysis (MC) \cite{shovkovy}, where the light quark chiral condensate was anticipated to rise. IMC is known to be driven by the competition between valence quark effects and sea quark leading to complex dynamics in the thermomagnetic environment \cite{ding2022chiral}.

The screening mass of mesons, particularly those in the pseudoscalar channel, serves as a critical observable for studying these dynamics. The neutral pion $\pi^0$ acts as the Goldstone boson associated with the spontaneous breaking of chiral symmetry, and its screening mass provides valuable information about chiral symmetry restoration at finite temperature and magnetic field. By analyzing the behavior of screening masses, one can gain insight into the long-distance properties of the medium and the changes in mesonic characteristics near the QCD phase transition \cite{bazavov2019meson}.

This work focuses on the screening masses of neutral pseudoscalar mesons under finite temperature and magnetic field conditions. Using lattice QCD simulations, we investigate their dependence on temperature, magnetic field strength, and quark combination. The continuum estimate of the results are taken to provide a comprehensive understanding of the interplay between magnetic fields and QCD phase transitions. This proceeding is directly based on the work presented in Ref.~\cite{Ding:2025pbu}.

\section{Theory}
\subsection{Ward-Takahashi Identity}
The pseudoscalar meson properties are intricately linked to the quark chiral condensates via the Ward-Takahashi identities. These identities can be expressed as \cite{ding2021}:
\begin{align}
    (m_u + m_d)\chi_{\pi^0} &= \langle \bar{\psi}\psi \rangle_u + \langle \bar{\psi}\psi \rangle_d, \label{eq:WTI1} \\
    (m_d + m_s)\chi_{K^0} &= \langle \bar{\psi}\psi \rangle_d + \langle \bar{\psi}\psi \rangle_s, \label{eq:WTI2} \\
    m_s \chi_{\eta^0_{s\bar{s}}} &= \langle \bar{\psi}\psi \rangle_s, \label{eq:WTI3}
\end{align}
where $m_f$ and $\langle \bar{\psi}\psi \rangle_f$ represent the quark mass and chiral condensate for the flavors $f = u, d, s$. The susceptibilities $\chi_H$ correspond to neutral pseudoscalar mesons $H = \pi^0, K^0, \eta^0_{s\bar{s}}$, and are defined as the integrated two-point correlation functions:
\begin{equation}
    \chi_H(B,T)\equiv\int\text{d}z\int_{0}^{1/T}\text{d}\tau\int\text{d}x\int\text{d}y\,\mathcal{G}_H(B,\boldsymbol{x}),
\end{equation}
where $\boldsymbol{x} = (\tau, x, y, z)$, $B$ is the magnetic field strength, $T$ is the temperature, and $\mathcal{G}_H(B,\boldsymbol{x})$ is the meson correlation function for meson $H$, defined as $\mathcal{G}_{f_1,f_2}(B,\boldsymbol{x})=\mathcal{O}_{f_1,f_2}(B,\boldsymbol{x})\mathcal{O}^\dagger_{f_1,f_2}(B,0)$ and is obtained from the meson interpolator $\mathcal{O}_{f_1,f_2}(B,\boldsymbol{x})=\bar{\psi}_{f_1}(B,\boldsymbol{x})\gamma_5\psi_{f_2}(B,\boldsymbol{x})$.

\subsection{Screening Mass Formalism}
The screening mass $M_H$ is derived from the spatial correlator $G_H(B, T, z)$, which is obtained by projecting the meson correlator over zero momentum in the $x$, $y$, and $\tau$ directions:
\begin{equation}
    G_H(B, T, z) = \int_{0}^{1/T}\text{d}\tau\int\text{d}x\int\text{d}y\,\mathcal{G}_H(B,\boldsymbol{x}).
\end{equation}
At large spatial separations $z$, the correlator exhibits an exponential decay:
\begin{equation}
\underset{z \to \infty}{\lim} G_H(B,T,z)\sim A_H \exp(-M_H z),
\end{equation}
where $A_H$ is the amplitude and $M_H$ is the screening mass of meson $H$. On the lattice, periodic boundary conditions and finite spatial extent ensure symmetry of the screening correlator around $N_\sigma/2$ where $N_\sigma$ is the spatial extent of the lattice. For staggered mesons, the correlator oscillates due to the coupling of mesons of the same spin but opposite parities and is described by the ansatz: 
\begin{eqnarray}
    G_H(B, T, n_z) = \sum_{i}  A_H^{i, nosc} \cosh \big[M_H^{i, nosc}(N_\sigma/2 - n_z)\big] \nonumber\\
    - (-1)^{n_z} \sum_{j} A_H^{j, osc} \cosh \big[M_H^{j, osc}(N_\sigma/2 - n_z)\big],
\end{eqnarray} 
where $n_z$ is the spatial separation in lattice units,  and the indices $nosc$ and $osc$ denote the non-oscillating and oscillating contributions. 

\subsection{Relation with Chiral Symmetry}

The screening mass of a hadron is deeply connected to the chiral properties of the QCD medium. The light quark chiral condensate acts as an order parameter for the chiral symmetry. The Ward-Takahashi identities establish a direct link between the chiral condensate and the susceptibilities of neutral pseudoscalar mesons. These susceptibilities, in turn, are related to the screening mass via the spatial correlation functions of mesons. Thus, the screening mass provides complementary information about chiral symmetry. Unlike susceptibilities and chiral condensates, which are dominated by short-distance contributions and reflect bulk properties, screening masses focus on the long-distance behavior of the medium. Additionally, since the screening mass is the inverse of the screening length, it determines the range of interactions within the medium.

The light pseudoscalar meson $\pi^0$, as the Goldstone boson of spontaneous chiral symmetry breaking, plays a pivotal role in the context of chiral symmetry. At zero temperature, the pion screening mass equals the pole mass of the ground state. As the system approaches the chiral phase transition, the pion screening mass deviates from its zero-temperature value, signaling the gradual restoration of chiral symmetry. This deviation is a critical indicator of the changing dynamics of the medium and highlights the influence of temperature on mesonic properties. 

In the presence of a magnetic field, the behavior of the screening mass becomes more intricate due to explicit symmetry breaking. The QCD chiral symmetry transitions from $SU_L(2) \times SU_R(2) \sim O(4)$ to $U_L(1) \times U_R(1) \sim O(2)$ in two-flavor QCD. This explicit breaking causes the charged pions to acquire greater masses, leaving the neutral pion as the sole Goldstone boson. Consequently, the magnetic field significantly modifies the meson screening mass, offering valuable insights into the interplay between magnetic fields and chiral symmetry in the QCD medium.

\section{Lattice Simulation}
\subsection{Setup}
The lattice QCD simulations in this study are performed with $(2+1)$-flavor QCD using the Highly Improved Staggered Quark (HISQ/tree) action and a tree-level improved Symanzik gauge action. The strange quark mass $m_s$ is tuned to its physical value by matching the mass of the fictitious $\eta_{s\bar{s}^0}$ meson to $M_{\eta_{s\bar{s}}^0} = \sqrt{2M_K^2 - M_\pi^2} \approx 684$ MeV obtained using the chiral perturbation theory. The up and down quarks have equal masses $m_l$ and are given by $m_l=m_s/27$. The lattice temperature is varied near the pseudocritical temperature $T_{pc}$ with five temperature values studied for each lattice dimension close to $T_{pc}$.

The pion mass is set to its physical value, $M_\pi= 135$ MeV, and the kaon decay constant $f_K$ is used to set the lattice scale \cite{bazavov2019meson}. The aspect ratio of the lattice is fixed, $N_\sigma / N_\tau = 4$. Simulations are performed with $N_\tau = 8$, 12, and 16 to allow for studying the effects of varying lattice resolutions.

\subsection{Magnetic Field Implementation}
An external magnetic field is introduced along the $z$-direction. To maintain periodic boundary conditions, the magnetic flux is quantized, satisfying the relation \cite{DElia:2010abb,bali2012qcd}:
\[
eB = \frac{6\pi N_b}{N_\sigma^2 a^2} = \frac{6\pi N_b T^2 N_\tau^2}{N_\sigma^2},
\]
where $N_b \in \mathbb{Z}$ is the quantized flux, $a$ is the lattice spacing, and $T$ is the temperature. This quantization ensures that the magnetic field is consistent with the lattice periodicity. Nine values of $N_b = 0, 1, 2, 3, 4, 6, 12, 16$ and $24$ corresponding to $eB$ ranging from $0$ to $1$~GeV$^2$ for each temperature are used in this study.

\subsection{Simulation Details}
All gauge configurations are generated using a modified version of the SIMULATeQCD code \cite{mazur2024}, which incorporates the magnetic field effects, with configurations saved at every tenth-time step. Corner-wall sources \cite{gupta2003} are employed for correlation function calculations. Single corner-wall sources are used for lattices with $N_\tau = 8$ and 12, while four corner-wall sources are utilized for $N_\tau = 16$ to reduce noise. The contribution of the disconnected quark lines to the screening mass is neglected in this analysis, as their impact is expected to be small \cite{ding2021}.

\section{Analysis Methodology}

\subsection{Correlation Function and Screening Mass Extraction}

\begin{figure*}
\centering
\includegraphics[scale=0.4]{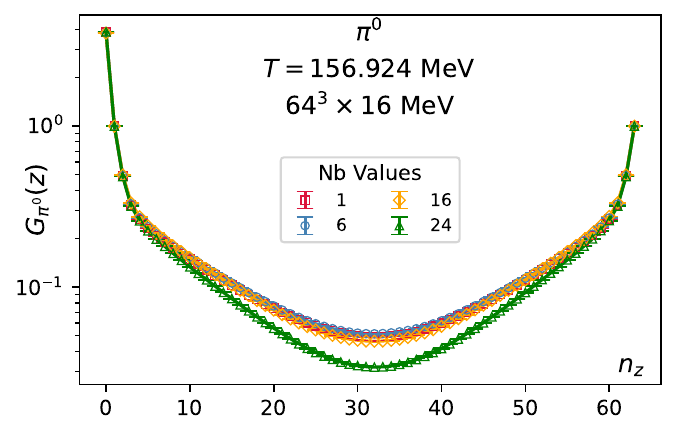}
\includegraphics[scale=0.5]{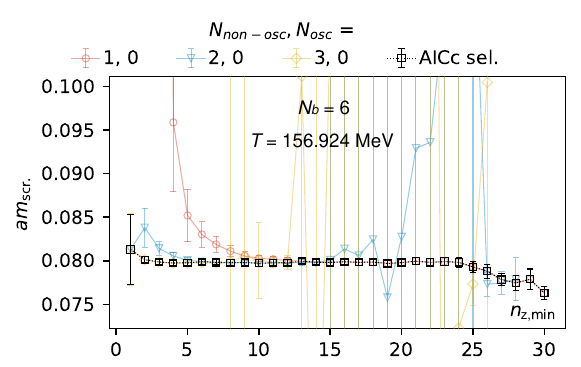}
\includegraphics[scale=0.5]{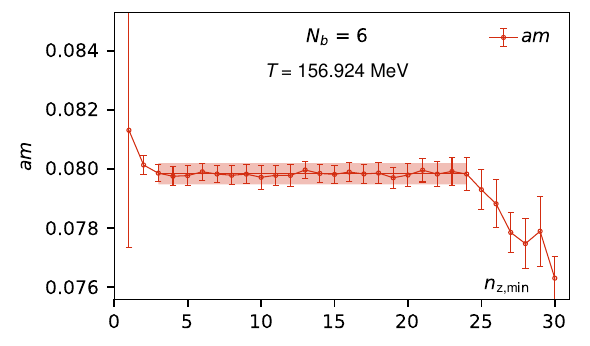}
\caption{Sample screening correlators (left), AICc selection (middle) and screening mass plateau (right) at $T=156.924$ MeV on a lattice with size $64^3\times 16$ for $\pi^0$. The correlator is shown for four values of magnetic flux $N_b$ while the AICc selection and plateau is shown for $N_b=6$.}
\label{fig:correlation}
\end{figure*}

Sample normalized screening correlators for the neutral pion \(\pi^0\)meson are shown in ~\autoref{fig:correlation} (left), corresponding to lattice sizes \(N_s = 64\) for four $N_b$ values at $T=156.924$ MeV. The correlator is normalized at $n_z=1$.  

To extract the screening masses, we fit the correlators using a range of ansatz configurations that account for both non-oscillating and oscillating contributions \cite{bazavov2019meson}. Up to three states were considered in both non-oscillating and oscillating terms to obtain fit parameters. Fits were performed over the folded correlator to decrease reduce the fitting error and over the interval \([n_{z,\text{min}}, N_z / 2]\), where \(n_{z,\text{min}}\) ranges from 0 to \(N_z / 2 - 2\). 

Model selection for the best fit was achieved using the corrected Akaike Information Criterion (AICc):
\begin{equation}
\text{AICc} = 2k - \ln(\hat{L}) + \frac{2k^2 + 2k}{n - k - 1},
\end{equation}
where \(k\) is the number of parameters, \(\hat{L}\) is the likelihood function, and \(n\) represents the number of data points. The AICc criterion effectively penalizes models with excessive parameters, preventing overfitting. In most cases, the non-oscillating ansatz was chosen as the best representation of the data. ~\autoref{fig:correlation} (middle) shows the masses obtained for different ansatz considered along with the AICc selection for $N_b=6$ at $T=156.924$ MeV with lattice size $64^3\times 16$ for $\pi^0$.

Screening masses and their uncertainties were determined by identifying a plateau in the AICc-selected results, with final central estimate values being the median and the error estimated as 68.27\% quantile extracted using a Gaussian bootstrapping method. ~\autoref{fig:correlation} (right) shows the mass plateau obtained from AICc selection for $N_b=6$ at $T=156.924$ MeV with lattice size $64^3\times 16$ for $\pi^0$.

\subsection{Interpolation in $T-eB$ plane}
\begin{figure*}
\includegraphics[scale=0.41]{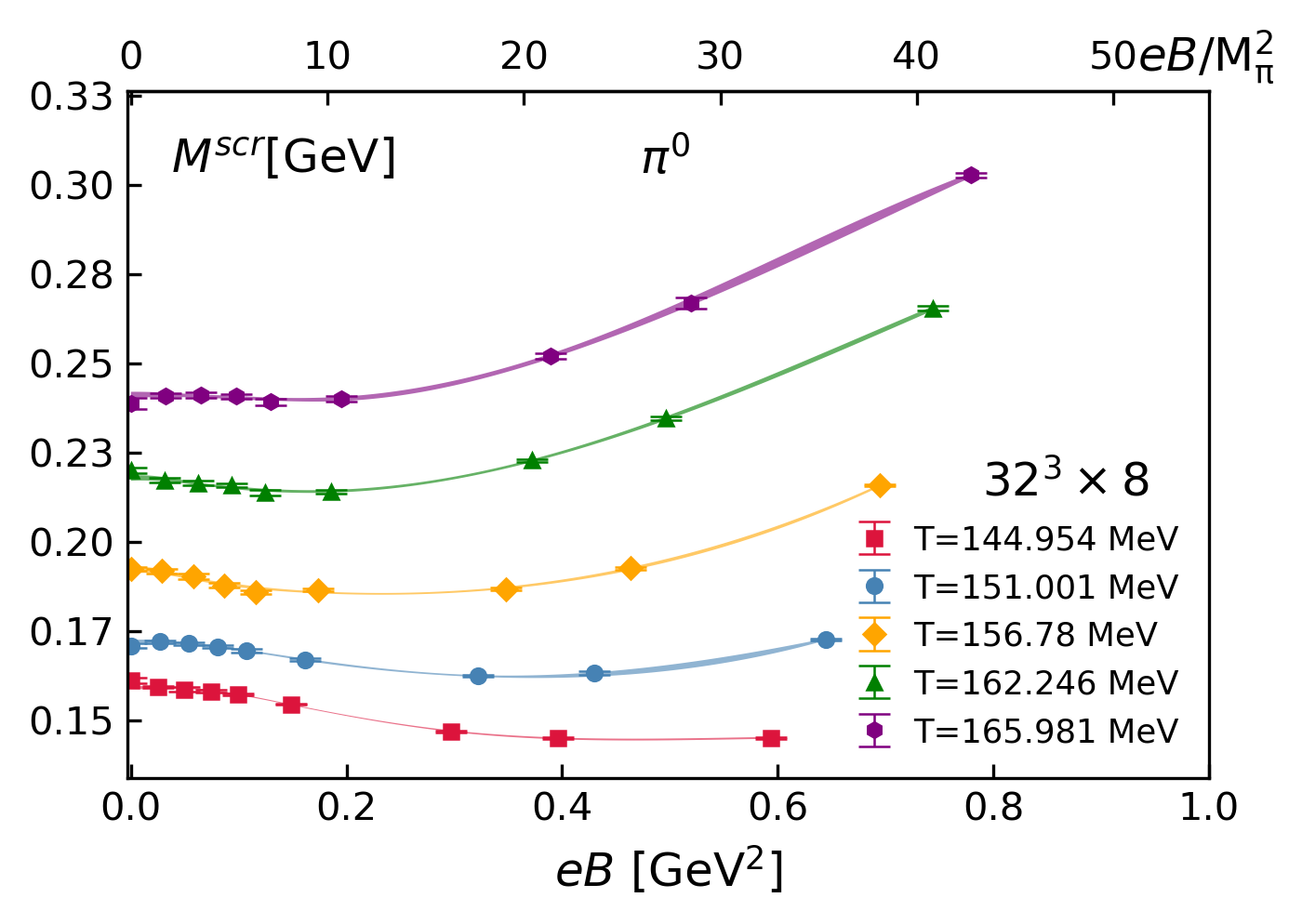}
\includegraphics[scale=0.41]{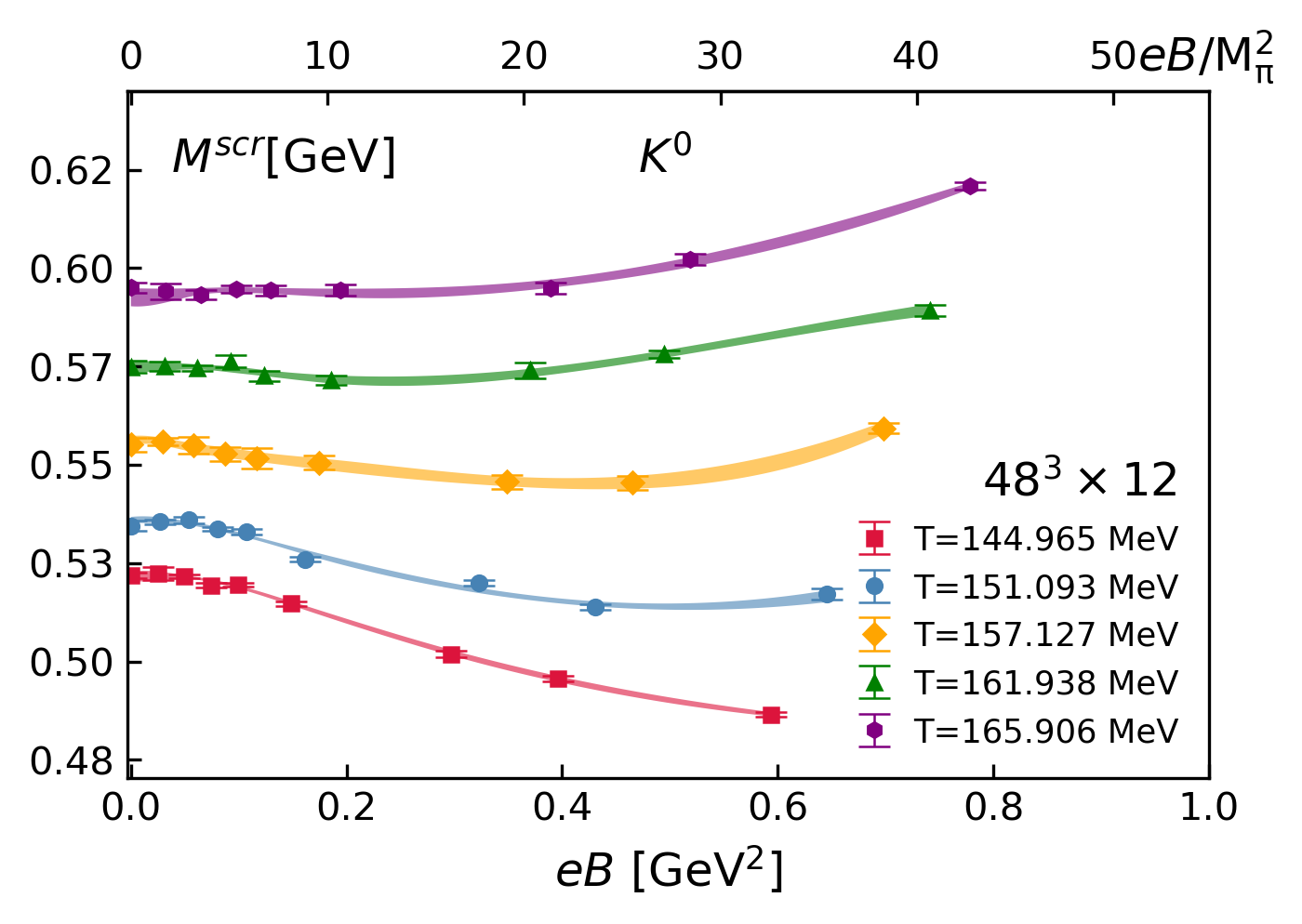}
\includegraphics[scale=0.41]{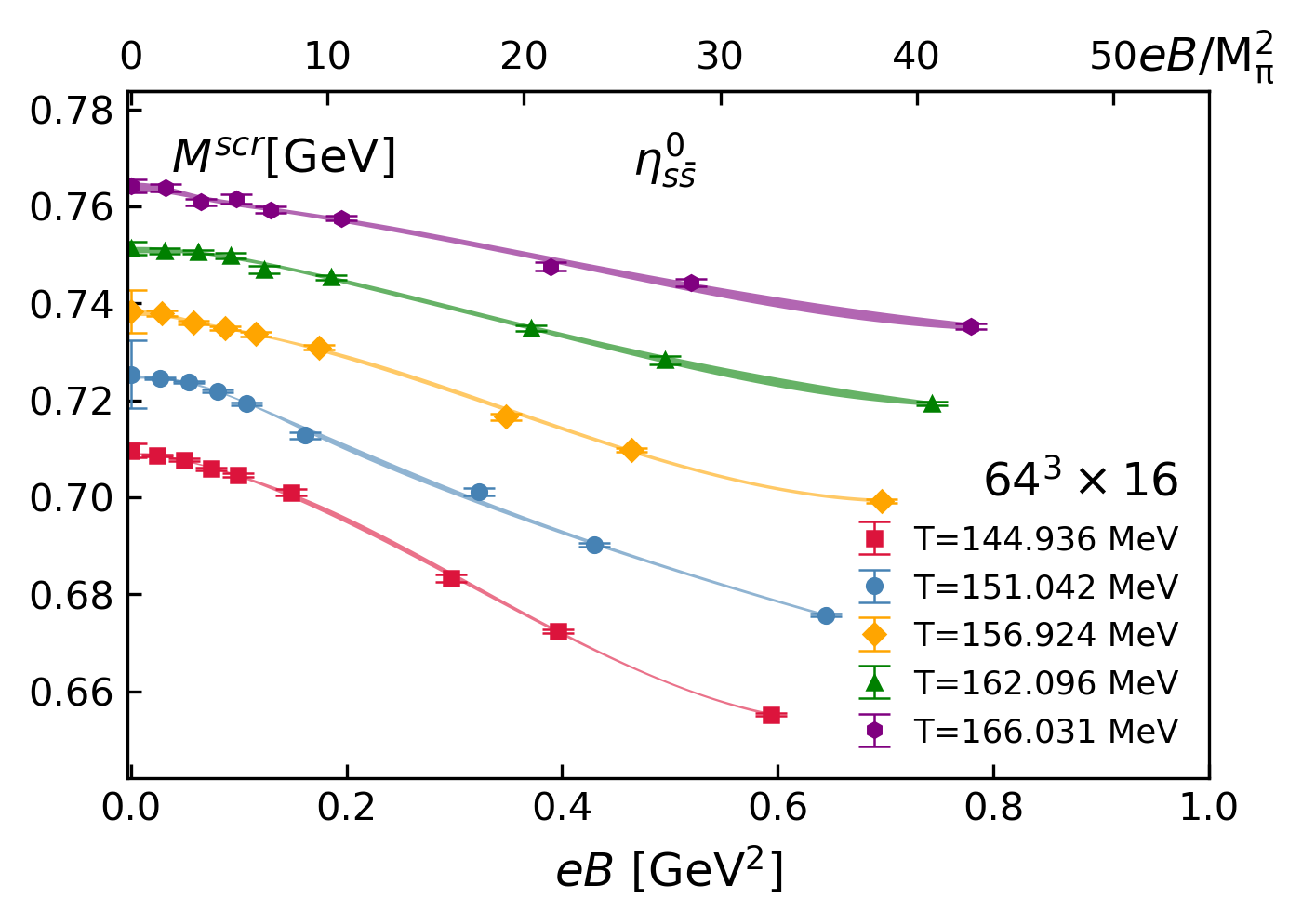}
\caption{ Samples of interpolation estimate of the screening mass as a function of the magnetic field
$eB$ at fixed temperatures are shown for three quark-lattice dimension combinations:  $32^3\times 8$ for $\pi^0$ (left), $48^3\times 12 $ for $K^0$ (middle), $64^3\times 16$ for $\eta^0_{\bar{s}s}$ (right). The data points are the lattice data and the band depicts the interpolated values in $T - eB$ plane. The upper x-axis is rescaled by the pion mass square in the vacuum at eB = 0 to make it dimensionless.}
\label{interpolation}
\end{figure*}

To estimate the screening masses at intermediate points within the \(T\)-\(eB\) plane, a 2-dimensional B-spline interpolation method was employed using the  "bisplrep" function of the scipy package. The algorithm automatically determines the number and positions of the interpolation knots based on the data, while a smoothing factor is applied to regulate the interpolation process. The uncertainty bands for the interpolated values were calculated using a bootstrap resampling technique: multiple samples were generated from the original data, the interpolation was performed on each sample, and the central value was taken as the median of these results. The error was then determined using the 68.27\% quantile.

~\autoref{interpolation} illustrates the sample interpolation results for the screening masses for different mesons and lattice size combination: \(32^3 \times 8\) for the neutral pion \(\pi^0\) on the left, \(48^3 \times 12\) for the neutral kaon \(K^0\) in the center, and \(64^3 \times 16\) for the \(\eta^0_{s\bar{s}}\) meson on the right. The data points represent results from direct lattice simulations, while the shaded regions display the interpolated trends.

\subsection{Continuum Estimate}
\begin{figure*}
{\includegraphics[scale=0.41]{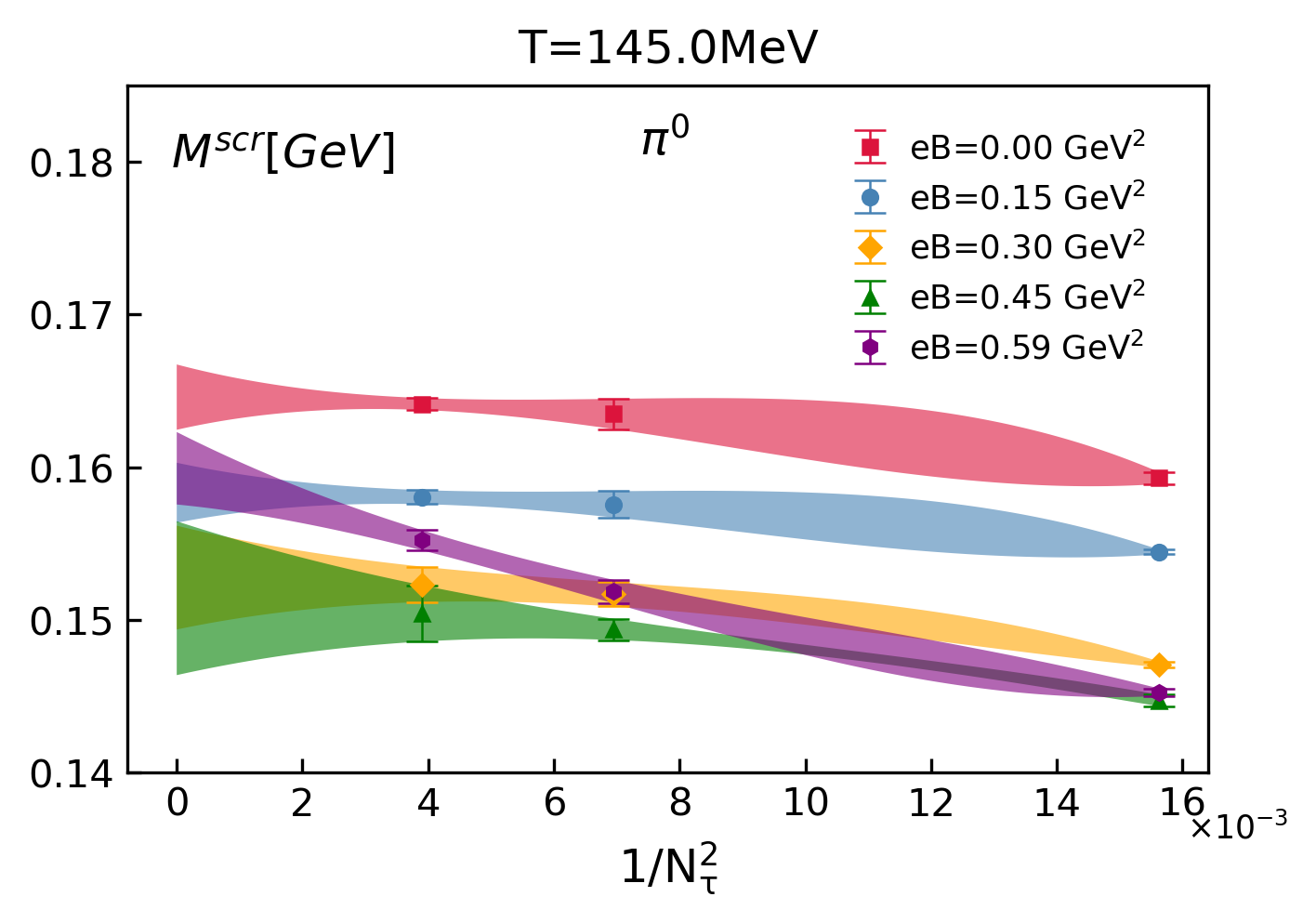}}
\includegraphics[scale=0.41]{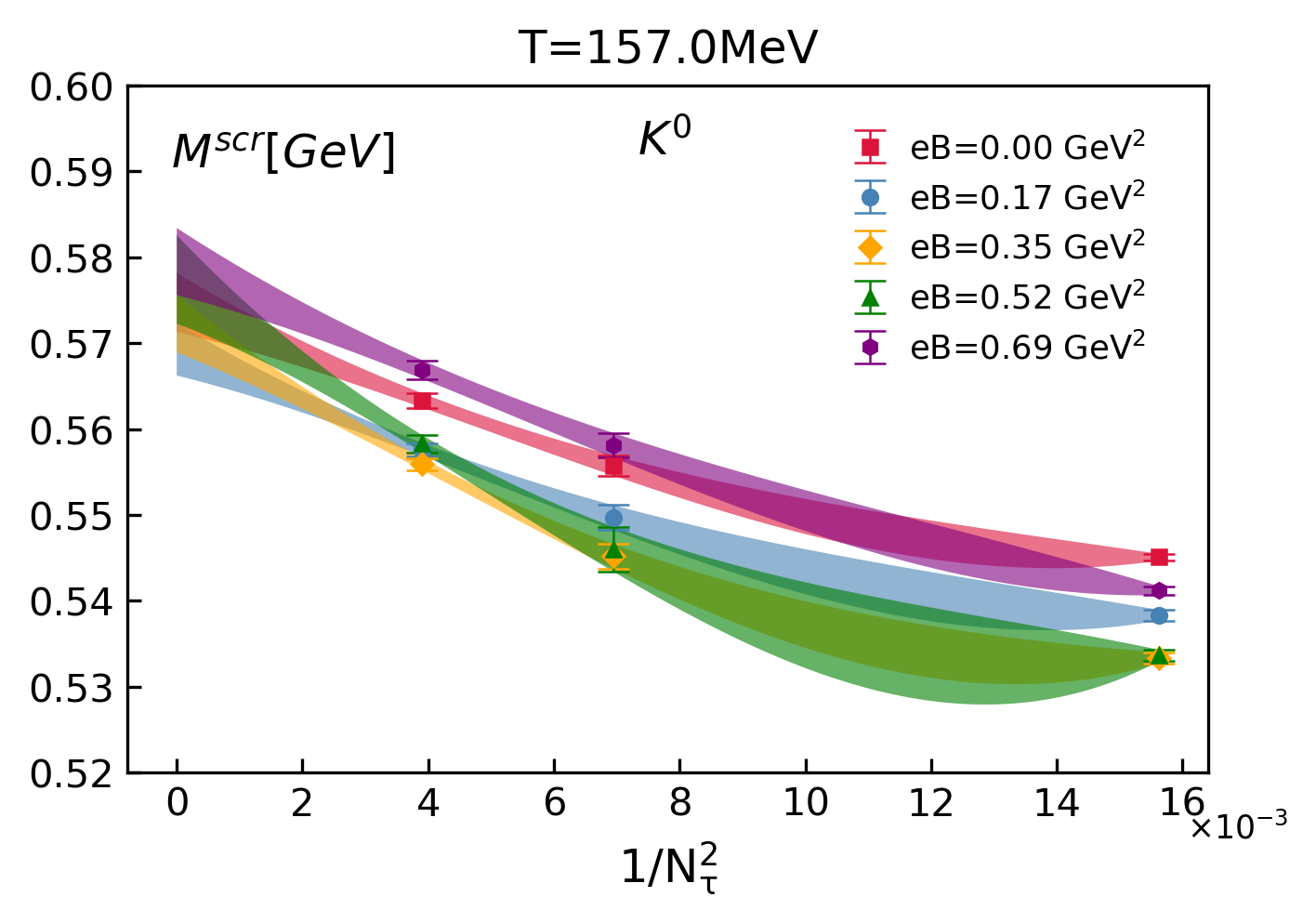}
\includegraphics[scale=0.41]{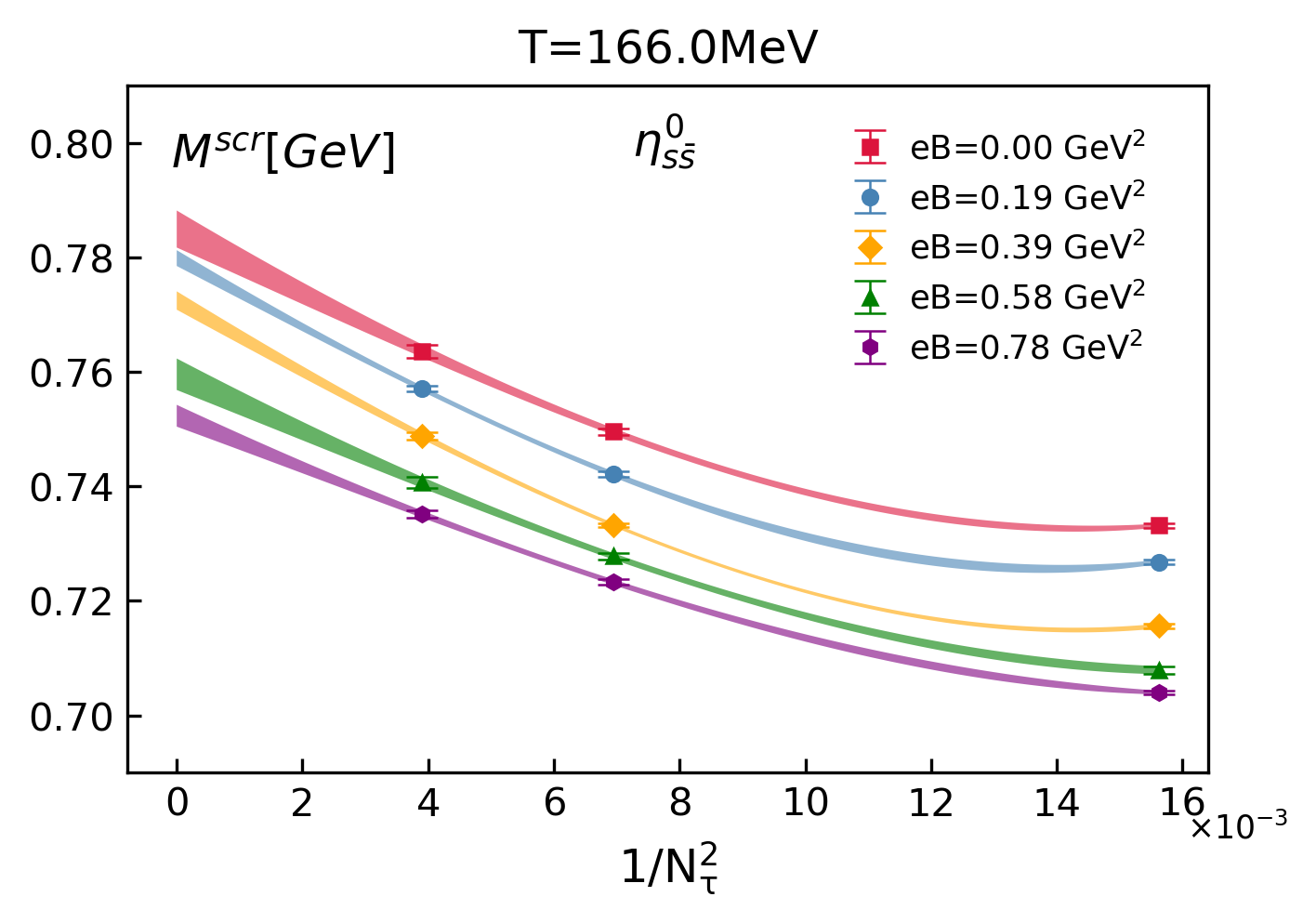}
\caption{ Samples of continuum estimation of the screening mass at fixed temperatures $T$ and magnetic field $eB$ are shown for three temperatures and quark combination: $\pi^0$ at T=145.0 MeV (left), $K^0$ at T=157.0 MeV (middle), and $\eta^0_{\bar{s}s}$ at T=166.0 MeV (right). The data points are obtained through interpolation of the lattice data and the band depicts the error obtained using the RMS of linear and quadratic ansatz.}
\label{estimation}
\end{figure*}
For obtaining the continuum estimate, we expanded the interpolated screening mass in terms of \(1/N_\tau^2\) to account for lattice spacing effects. Linear and quadratic fits of the form:
\begin{eqnarray}
O(T, B, N_\tau) = O_{\text{lin}}(T, B) + \frac{b}{N_\tau^2}\\
O(T, B, N_\tau) = O_{\text{quad}}(T, B) + \frac{c}{N_\tau^2} + \frac{d}{N_\tau^4},
\end{eqnarray}
were combined to minimize systematic uncertainties. Here, $O$ is the observable, and $b,c,$ and $d$ are fit constants. The linear ansatz is applied to lattices with $N_\tau = 12$ and 16, while quadratic ansatz is applied to lattices with $N_\tau = 8, 12$ and 16.
The final continuum estimates were obtained by bootstrapping the averaged results of these fits, with median being the central estimate and errors estimated from the 68.27\% quantile.

Figure~\autoref{estimation} shows examples of the continuum estimates of screening masses at fixed temperatures \(T\) and magnetic field strengths \(eB\). The results are displayed for three neutral pseudoscalar mesons at specific representative temperatures: \(\pi^0\) at \(T = 145.0\) MeV (left), \(K^0\) at \(T = 157.0\) MeV (middle), and \(\eta^0_{\bar{s}s}\) at \(T = 166.0\) MeV (right). The data points represent values obtained from the interpolation of lattice data, while the shaded bands indicate the uncertainties estimated using the root mean square (RMS) of the linear and quadratic ansatz fits.

\section{Results}
\subsection{Screening Mass vs Magnetic Field Strength}
\begin{figure*}
\includegraphics[scale=0.42]{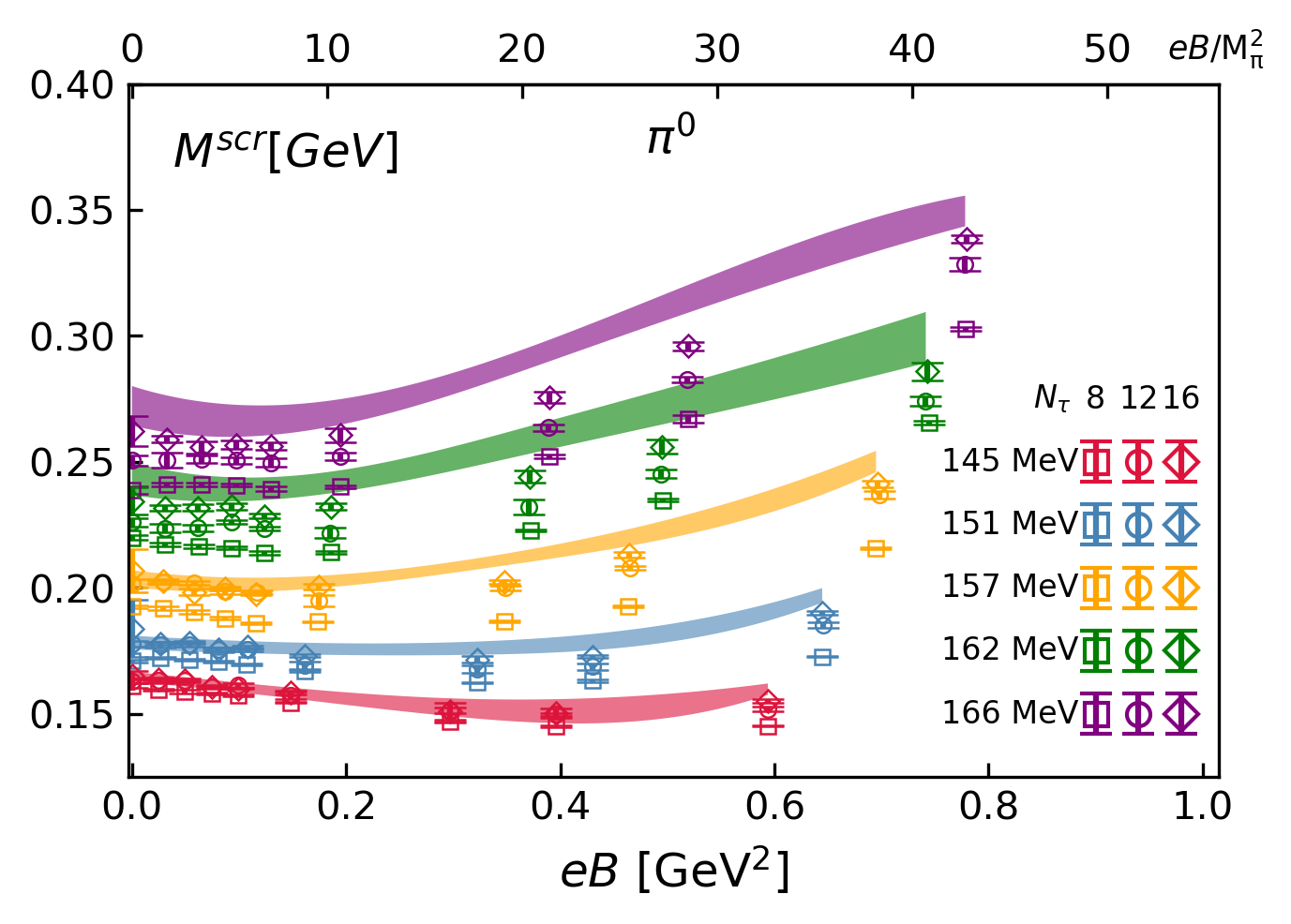}
\includegraphics[scale=0.42]{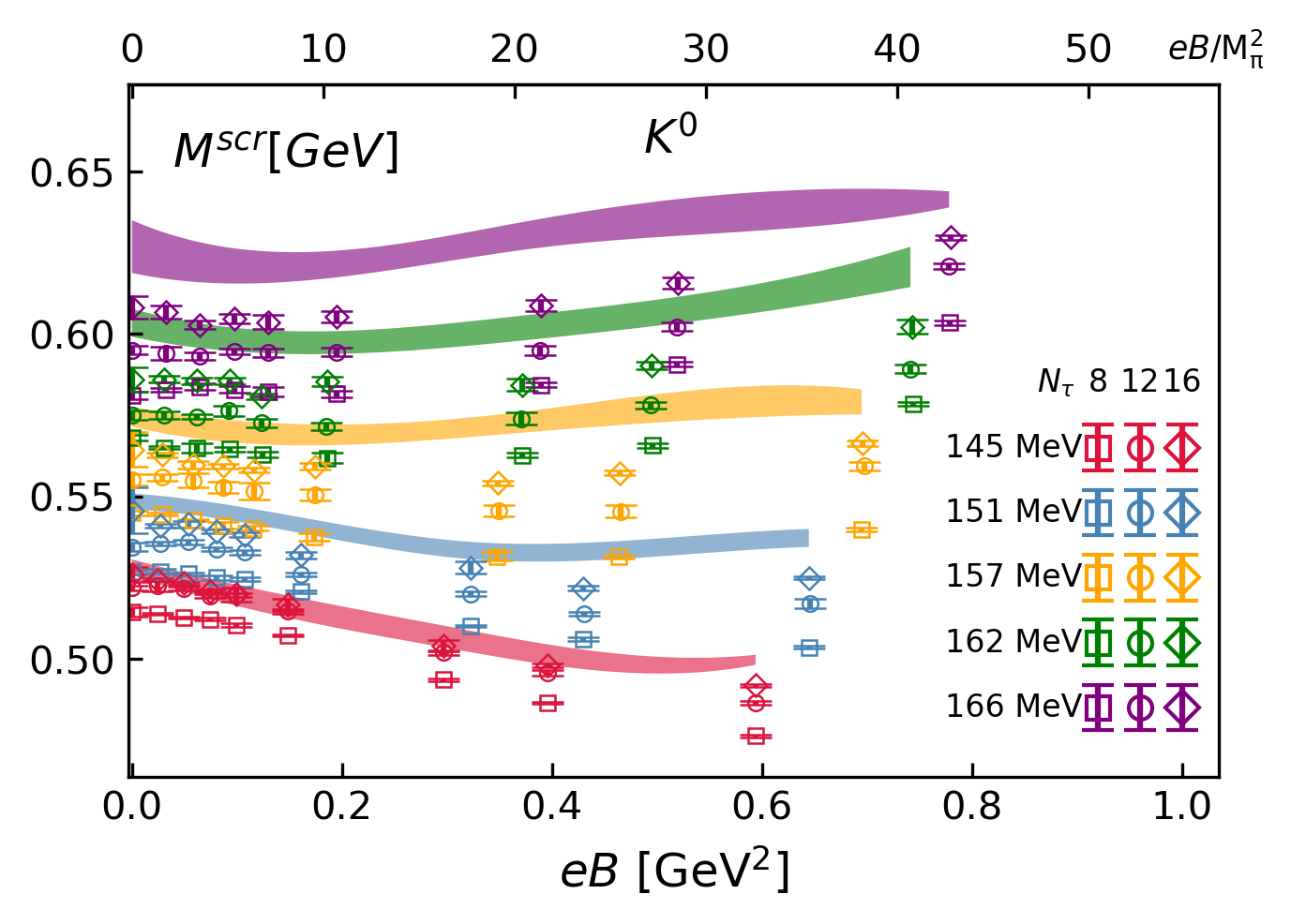}
\includegraphics[scale=0.42]{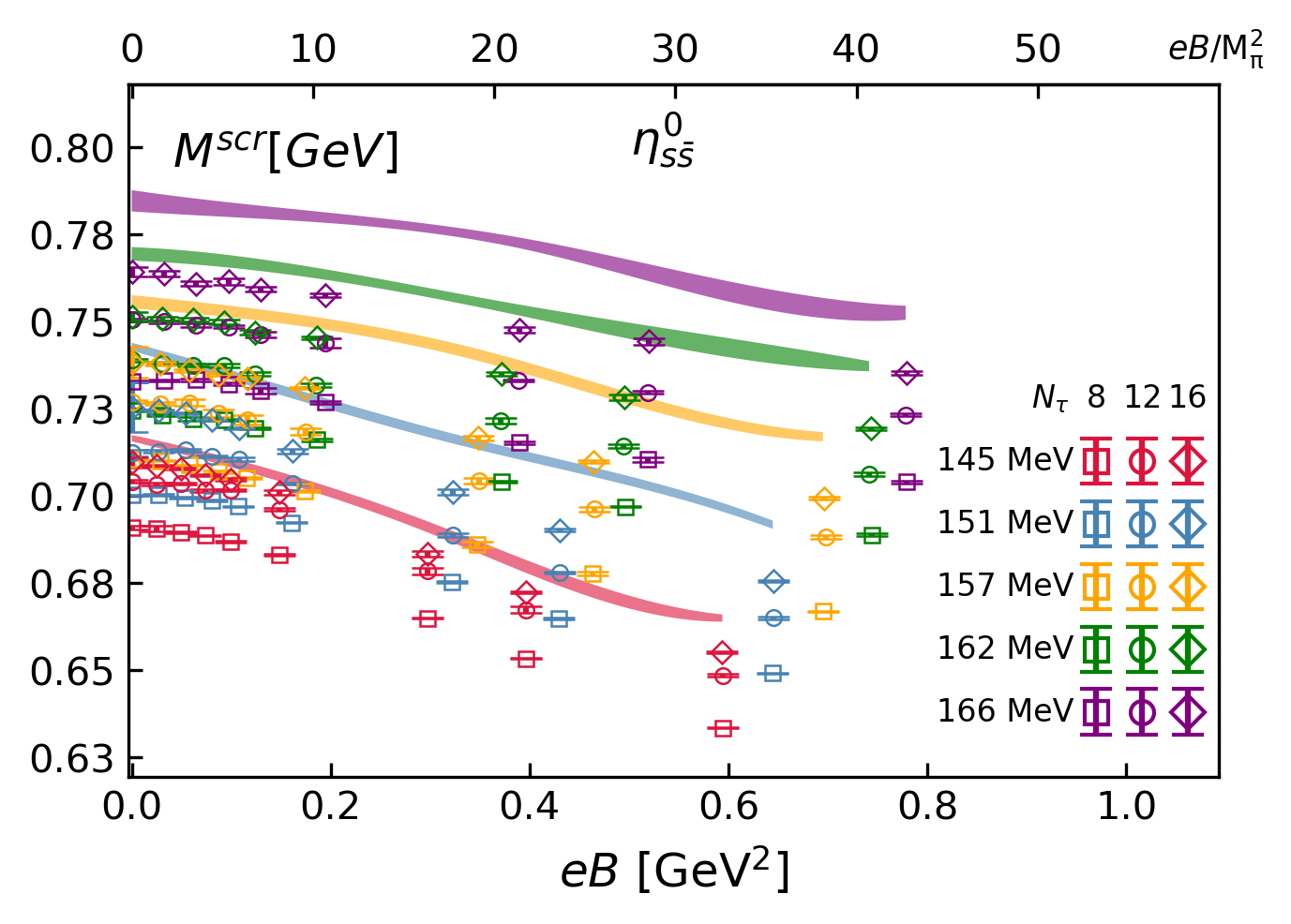}
\caption{The continuum estimates of the screening mass for neutral pseudoscalar mesons, namely $\pi^0$ (left), $K^0$ (middle), $\eta^0_{\bar{s}s}$ (right) as a function of magnetic field strength $eB$ at a fixed temperature
$T$. The shaded bands represent the continuum estimated results, while the data points correspond to lattice data at temperatures rounded to the nearest integer. The upper x-axis is rescaled by the pion mass square in the vacuum at $eB = 0$ to make it dimensionless.}
\label{mscrn_vs_eb}
\end{figure*}
In figure~\autoref{mscrn_vs_eb}, we present the dependence of screening masses on the magnetic field strength \(eB\) for the neutral pion \(\pi^0\), kaon \(K^0\), and the fictitious eta meson \(\eta^0_{\bar{s}s}\) at various fixed temperatures near the pseudocritical temperature \(T_{pc}\).

For the neutral pion \(\pi^0\), the screening mass initially decreases as \(eB\) increases, reaching a minimum before rising again at larger values of \(eB\). As the temperature increases, the position of the minimum shifts to smaller values of \(eB\). This phenomenon can be attributed to the inverse chiral magnetic effect induced by sea quarks, as discussed in prior work with heavier quark masses \cite{ding2022chiral}. 

The neutral kaon \(K^0\) exhibits a similar trend, but the minima are more pronounced, particularly at lower temperatures. Additionally, at higher \(eB\), the increase in screening mass is less steep for \(K^0\) compared to \(\pi^0\), suggesting a gentler slope for the kaon in strong magnetic fields.

In contrast, the behavior of the fictitious eta meson \(\eta^0_{\bar{s}s}\) differs significantly. The screening mass decreases steadily with increasing \(eB\), with a more pronounced slope observed at lower temperatures. Unlike \(\pi^0\) and \(K^0\), no minima are observed for \(\eta^0_{\bar{s}s}\), highlighting the dominance of valence quarks and magnetic catalysis effect observed for the strange quark.

\subsection{Screening Mass vs Temperature}

\begin{figure}
\includegraphics[scale=0.42]{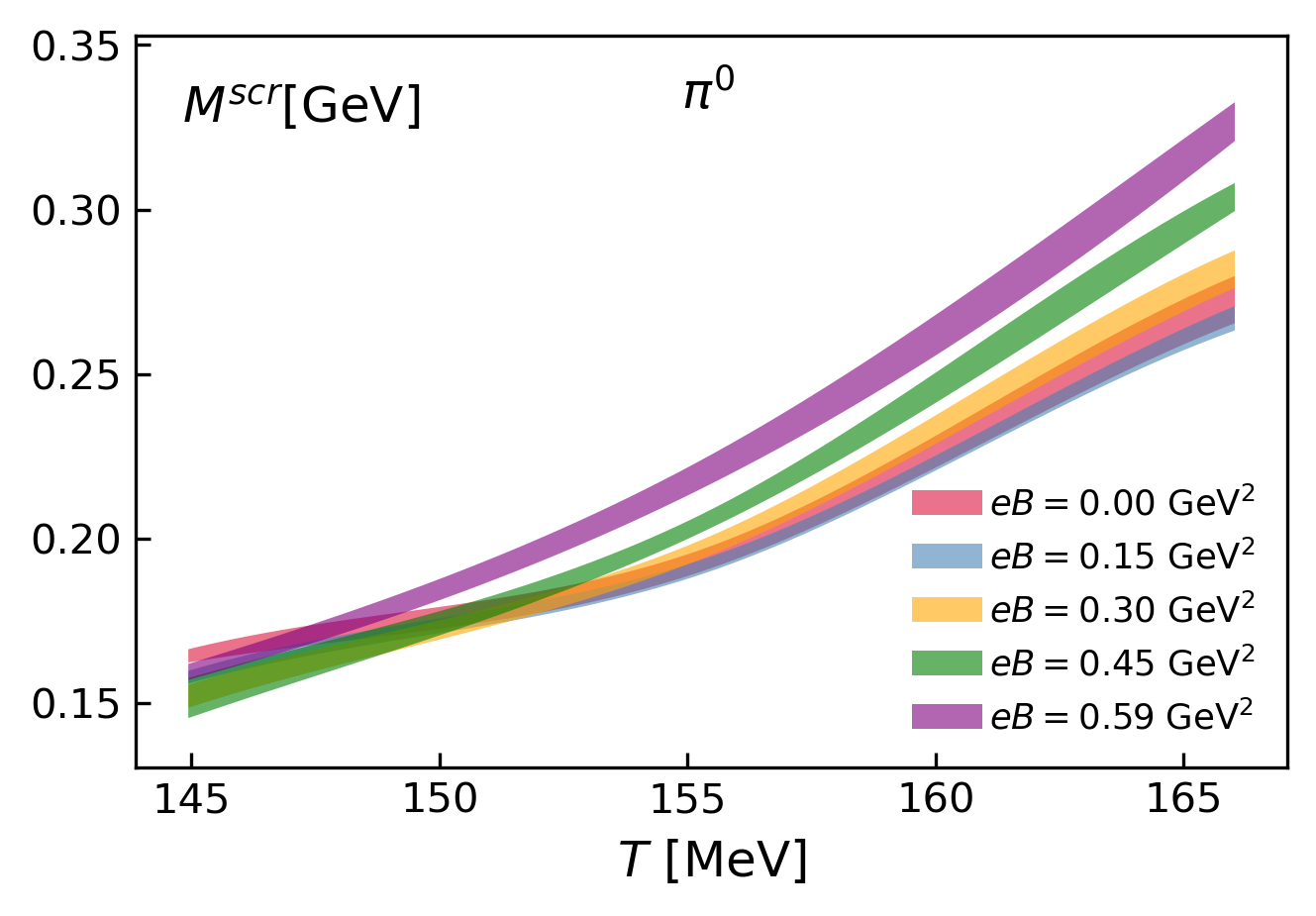}
\includegraphics[scale=0.42]{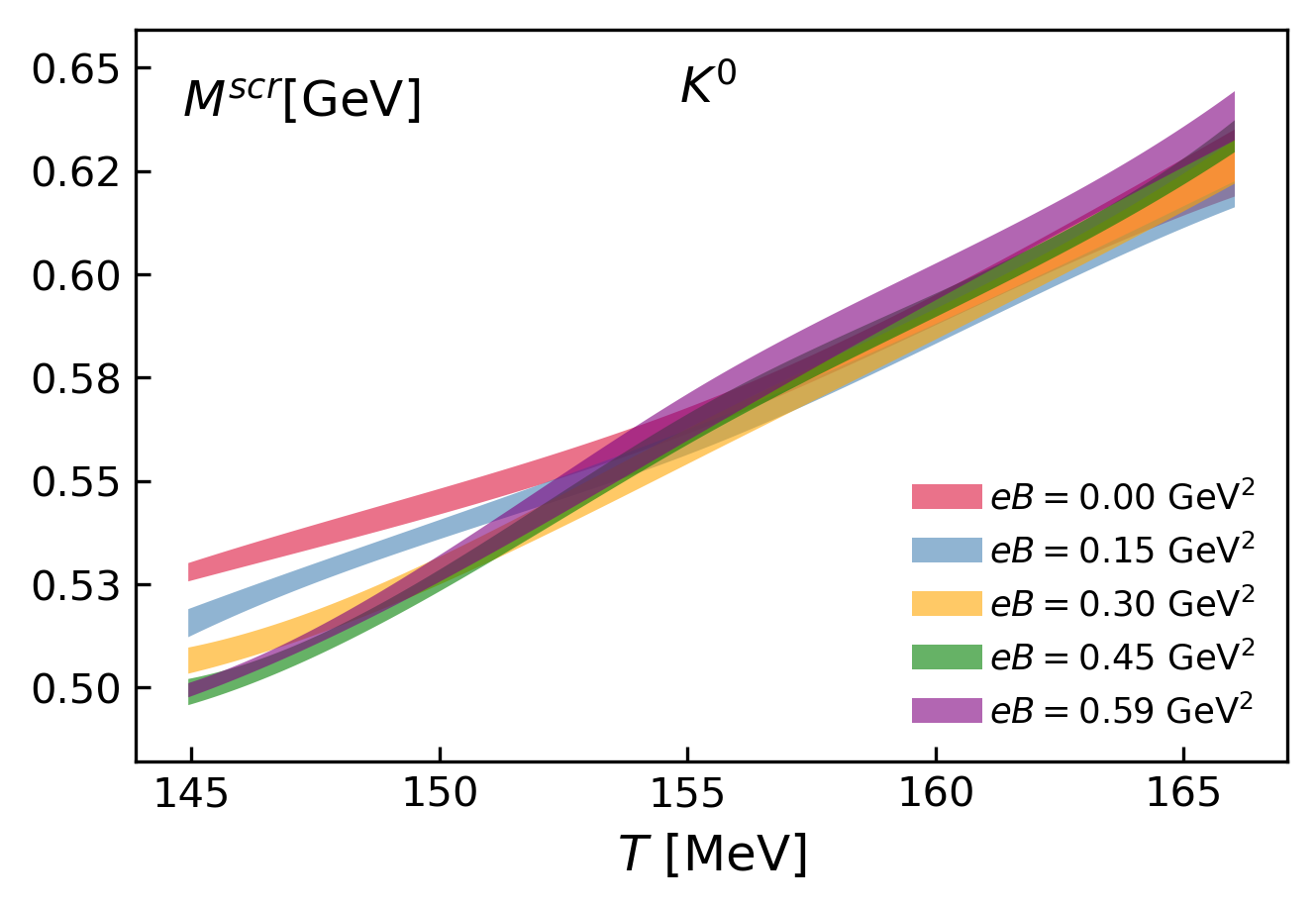}
\includegraphics[scale=0.42]{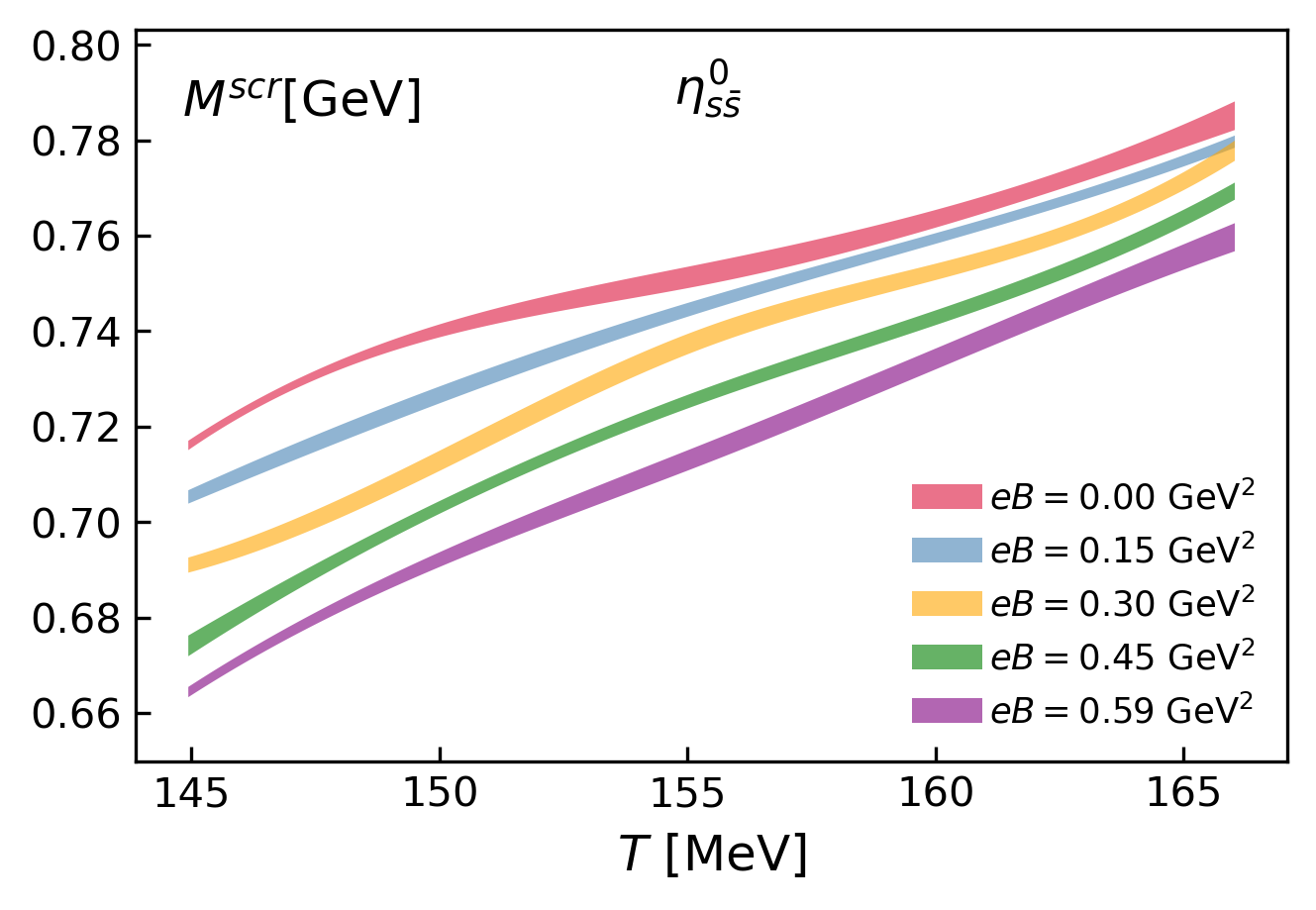}
\caption{ The continuum estimates of the screening mass for neutral pseudoscalar mesons, namely $\pi^0$ (top), $K^0$ (middle), $\eta^0_{\bar{s}s}$ (bottom) as a function of temperature
T at fixed magnetic field eB. The band depicts the continuum estimates.}
\label{mscrn_vs_T}
\end{figure}

Figure~\autoref{mscrn_vs_T} shows the screening masses as functions of temperature for fixed values of \(eB\). The panels are organized similarly to Fig.~\autoref{mscrn_vs_eb}, with results for \(\pi^0\), \(K^0\), and \(\eta_{s\bar{s}}^0\) shown from left to right.

For the \(\pi^0\) meson, the screening masses increase monotonically with temperature across all values of \(eB\) with the screening mass rising more steeply at higher $eB$. The constant temperature curves diverge at higher temperature, while at low temperatures, these curves intersect each other. This crossing behavior is consistent with a reduction in the pseudocritical temperature \(T_{pc}\) in the presence of stronger magnetic fields, a hallmark of inverse magnetic catalysis. This crossing behavior suggests a reduction in the pseudocritical temperature $T_{pc}$ in stronger magnetic fields. For the \(K^0\) meson, a similar crossing behavior is observed, although it occurs at slightly higher temperatures compared to the \(\pi^0\). 

For the \(\eta_{s\bar{s}}^0\) meson, the screening masses also increase with temperature. However, unlike the \(\pi^0\) and \(K^0\), no crossing behavior is observed, and the curves appear to converge at higher temperatures.

\section{Conclusion}

In this study, we investigated the screening masses of neutral pseudoscalar mesons in a thermomagnetic QCD medium near the pseudocritical temperature \(T_{pc}\) using continuum-estimated results of lattice QCD simulations. The neutral pion (\(\pi^0\)) and kaon (\(K^0\)) mesons exhibit convex non-monotonic behavior in their screening masses, encountering minima when turning on the magnetic field, reflecting a delicate balance between the inverse magnetic catalysis (IMC) and magnetic catalysis (MC) effects. In contrast, the \(\eta^0_{\bar{s}s}\) meson displays a continuous decrease in its screening mass, demonstrating the dominant MC effect on mesons with strange quark.

The temperature dependence of the screening masses further emphasizes the impact of IMC, with the reduction of \(T_{pc}\) manifested through the crossing behavior of the \(\pi^0\) and \(K^0\) screening masses at higher \(eB\). This behavior highlights how stronger magnetic fields promote earlier chiral symmetry restoration, consistent with the observed suppression of the chiral condensate in the presence of the magnetic field. The interplay between sea and valence quarks in this thermomagnetic environment is crucial for understanding mesonic properties.

%\section*{Acknowledgments}
\acknowledgments
This work was supported in part by the National Natural Science Foundation of China under Grants No. 12293064, No. 12293060, and No. 12325508, and the National Key Research and Development Program of China under Contract No. 2022YFA1604900. The numerical simulations were performed using the GPU cluster at the Nuclear Science Computing Center of Central China Normal University (NSC3) and the Wuhan Supercomputing Center.

\end{document}